\documentclass[prb,twocolumn,superscriptaddress,showpacs,nofootinbib]{revtex4-1}

\usepackage{amsxtra}			% for boldsymbol
\usepackage[percent]{overpic} 	% for figure overlays

\usepackage{tabularx}

\usepackage[
pdfpagelabels, 
bookmarks = true,
plainpages=false,
bookmarksopen = true,
bookmarksnumbered = true,
breaklinks = true,
hypertexnames=true,
hidelinks,
linkcolor = black,
urlcolor  = red,
citecolor = black,
anchorcolor = black,
unicode,
hyperindex = true,
hyperfigures,
]{hyperref}

\newcommand{\eq}[1]{Eq.~\ref{#1}}
\newcommand{\fig}[1]{Fig.~\ref{#1}}
\newcommand{\refr}[1]{Ref.~\onlinecite{#1}}
\newcommand{\app}[1]{Appendix~\ref{#1}}

\newcommand{\nll}{n}

\newcolumntype{R}{>{\centering\arraybackslash}X}

\graphicspath{{figures}{./figures/}}

\begin{document}

%%%%%%  TITLE  %%%%%%%%%%%%%%%%%%%%%%%%%

\title{Composite fermion model for entanglement spectrum of fractional quantum Hall states}

\author{Simon C. Davenport}
\affiliation{T.C.M Group, Cavendish Laboratory, J.J. Thomson Avenue, Cambridge CB3 0HE, United Kingdom}
\author{Iv\'{a}n D. Rodr\'{i}guez}
\affiliation{Max-Planck-Institut f\"{u}r Quantenoptik, 85748 Garching, Germany}
\author{J.~K. Slingerland}
\affiliation{Department of Mathematical Physics, National University of Ireland, Maynooth, Ireland}
\affiliation{Dublin Institute for Advanced Studies, School of Theoretical Physics, 10 Burlington Rd, Dublin, Ireland}
\author{Steven H. Simon}
\affiliation{Rudolf Peierls Centre for Theoretical Physics, 1 Keble Road, Oxford, OX1 3NP, UK}

\date{\today}

%%%%%%  ABSTRACT  %%%%%%%%%%%%%%%%%%%%%%

\begin{abstract}

We show that the entanglement spectrum associated with a certain class of strongly correlated many-body states --- the wave functions proposed by Laughlin and Jain to describe the fractional quantum Hall effect --- can be very well described in terms of a simple model of non-interacting (or weakly interacting) composite fermions. 

\end{abstract}

\maketitle

%%%%%%  INTRODUCTION   %%%%%%%%%%%%%%%%%

Understanding of strongly correlated many-body quantum states has been significantly enhanced in recent years by the introduction of the \emph{entanglement spectrum} (ES), i.e. the spectrum of eigenvalues of the reduced density matrix of a subsystem. \cite{li2008} The scrutiny of the ES has been particularly fruitful in the study of fractional quantum Hall (FQH) wave functions --- the archetype for topological phases of matter resulting from strong interaction effects --- where this analysis often reveals a special underlying entanglement structure useful for identifying and classifying different topological phases. \cite{li2008,haque2007,zozulya2007,dubail2012a,sterdyniak2012,rodriguez2012} 

For FQH states in particular it has been established that there is a deep connection between the entanglement structure of bulk wave functions (as seen in their ES \cite{qi2012,swingle2012,dubail2012b}) and the structure of edge excitations. \cite{lopez1991,moore1991,wen1992,cappelli1993} Central to this connection so far have been arguments based on the common conformal field theory description of the bulk and edge physics, which is known to apply for the class of wave functions that can be constructed in terms of expansion functions known as ``conformal blocks''. The ES for this class of what we shall refer to as ``simple'' FQH states has been extensively studied using the machinery of conformal field theory. \cite{sterdyniak2012,dubail2012a,dubail2012b}

Largely separate from these developments, it has been well documented that the composite fermion model of the FQH effect can successfully account for many experimentally observed features of the effect, particularly in the lowest Landau level. \cite{jainbook} It does so by positing that they are due to the integer quantum Hall (IQH) effect of non-interacting (or weakly interacting) quasiparticles known as composite fermions. While composite fermion theory presents an appealingly simple physical picture to explain the FQH effect, it has proved to be very challenging to relate the many-body composite fermion wave functions to constructs in conformal field theory.  While such constructs do exist, \cite{hansson2007a,hansson2007b,bergholtz2008,cappelli2013} they are somewhat complicated, and this has so far limited the effectiveness of conformal field theory machinery in describing the ES of the Jain states. In this sense the Jain states are not ``simple'' FQH states. 

In this work we present a very different approach to constructing entanglement spectra that bypasses the potential difficulty associated with explicitly writing wave functions in terms of conformal blocks. We show that  the low-lying (highest weight) part of the ES of the Jain FQH states can be accurately described by a modified ES of non-interacting (or weakly interacting) composite fermions in filled Landau levels (in other words a modified ES of the IQH states \cite{peschel2003,rodriguez2009,rodriguez2010}). To demonstrate the effectiveness of our method, we shall focus on presenting results for two fundamental examples, namely the Laughlin state at filling factor $\nu=1/2$ for bosons (a simple FQH state that was previously studied using conformal field theory methods \cite{dubail2012b}) and the Jain state at $\nu=2/3$ for bosons (a non-simple FQH state). (The method applies equally well to fermionic FQH states.) Our results for the Laughlin case are in good agreement with the approach based on conformal field theory, but the real advantage of our method becomes evident for the Jain states, where the conformal field theory approach appears much more complicated, and has therefore not been worked out.

In the spirit of the original composite fermion model, \cite{jainbook} we start with the ES for the IQH states, \cite{rodriguez2009,rodriguez2010} treating it in a general framework for determining the ES of non-interacting systems. \cite{peschel2003} We shall briefly review its derivation here. For simplicity we shall consider spinless particles. Also, to remove any additional complications due to edge physics we consider a system without physical edges. A standard technique to achieve this is to solve the problem on the surface of sphere. \cite{haldane1983} Here we have an integer $\nll$ Landau level problem (where the filling factor is $\nu=\nll$) for $N$ particles on a sphere of radius $\sqrt{Q}$ (in units of magnetic length) that encloses a fictitious magnetic monopole of strength $2Q=(N-\nll^2)/\nll$. There is rotational symmetry about the $z$-axis, which leads to the single-particle orbitals of the problem being labelled by the z-component of angular momentum $m=-(2Q+\sigma)/2,-(2Q+\sigma)/2+1,\ldots,(2Q+\sigma)/2$ in addition to a Landau level index $\sigma=0,1,\ldots,\nll-1$. The single-particle orbitals are $\phi_{m,\sigma}(\mathbf r)$, where $\mathbf r $ lies on the surface of the sphere (technically, $\phi_{m,\sigma}(\mathbf r)$ are monopole harmonics. \cite{wu1976} The monopole harmonics also depend explicitly on $Q$, but for simplicity we suppress this dependence in our notation).

In order to connect to existing calculations we shall focus on the entanglement spectrum for wave functions in real space (as opposed to e.g. momentum space), which leads to the so-called real-space entanglement spectrum (RSES). \cite{sterdyniak2012,dubail2012a,rodriguez2012}

We shall consider cuts along lines of latitude, so that that rotational invariance about the $z$-axis is preserved and the $z$-component of angular momentum $L_{z}$ remains a good quantum number. Once the system is cut, $L_z$ is bipartitioned as $L_z = L^{A}_{z} + L^{B}_{z}$. Additionally, the total particle number is bipartitioned as $N = N_A + N_B$.

For the full fermionic $N$-particle system, we have a basis for the Hilbert space that consists of Slater determinants built from the orbitals $\phi_{m,\sigma}$. For the $A$ and $B$ subsystems, the restrictions of these orbitals to the regions $A$ and $B$ still form a complete single-particle basis, although the restricted orbitals must be renormalized to account for the fact that only part of the single-particle weight is in each subsystem. Once restricted to a subsystem, the orbitals $\phi_{m_1,\sigma_1}$ and $\phi_{m_2,\sigma_2}$ are orthogonal whenever $m_1\neq m_2$ (because the cut is made to preserve $L_{z}$), but are generally not orthogonal whenever $\sigma_1 \ne \sigma_2$ (though they remain linearly independent in this case). The Fock space of the $A$ and $B$ subspaces is spanned by Slater determinants in terms of these restricted orbitals. 

We can now write the Schmidt decomposition of the state $\left|{\psi}\right\rangle$ as 
\begin{equation}
\label{eqSchmidt}
\left|{\psi}\right\rangle=\sum_{i} e^{-\xi_{i}/2}\left|{\psi^{A}_{i}} \right\rangle \otimes \left|{\psi^{B}_{i}}\right\rangle
\end{equation}
where $\left|{\psi^{A}_{i}}\right\rangle$ and $\left|{\psi^{B}_{i}}\right\rangle$ belong to the Fock spaces for the $A$ and $B$ subsystems that we have just described. The non-zero Schmidt coefficients, $e^{-\xi_{i}/2}$, are conveniently written in terms of the entanglement energies, $\xi_{i}$. The index $i$ labels the vectors in the Schmidt basis for the $A$ system (at least those with non-zero Schmidt coefficients). Since $N_{A}$ and $L^{A}_{z}$ are conserved by the cut, we can choose the Schmidt basis to consist of eigenstates of $N_A$ and $L^{A}_{z}$. We can then write $\xi_{N_{A},L^{A}_{z},i}$ for the entanglement energy of the $i^{\rm th}$ Schmidt state with given $N_{A}$ and $L^{A}_{z}$.  In fact, with any Schmidt state with $N_A$ particles and given $L^{A}_{z}$, we can associate an $N_{A}$-tuple of single particle momenta $\mathbf{m}=(m_1,\ldots,m_{N_{A}})$ with $L^{A}_{z}=\sum_{p}m_{p}$ and $m_1 < m_2< \ldots < m_{N_{A}}$. If there is only a single Landau level involved, the Schmidt state is just the unique Slater determinant in the Fock space of system $A$ labelled by $\mathbf{m}$ and we can replace the label $i$ above by $\mathbf{m}$. With multiple Landau levels, there will be a number of Schmidt states associated with each $\mathbf{m}$. These states are superpositions of the Slater determinants with this $\mathbf{m}$ and different choices of the numbers of particles in each Landau level.

\begin{figure}[t]

\begin{overpic}[width=0.5\textwidth]{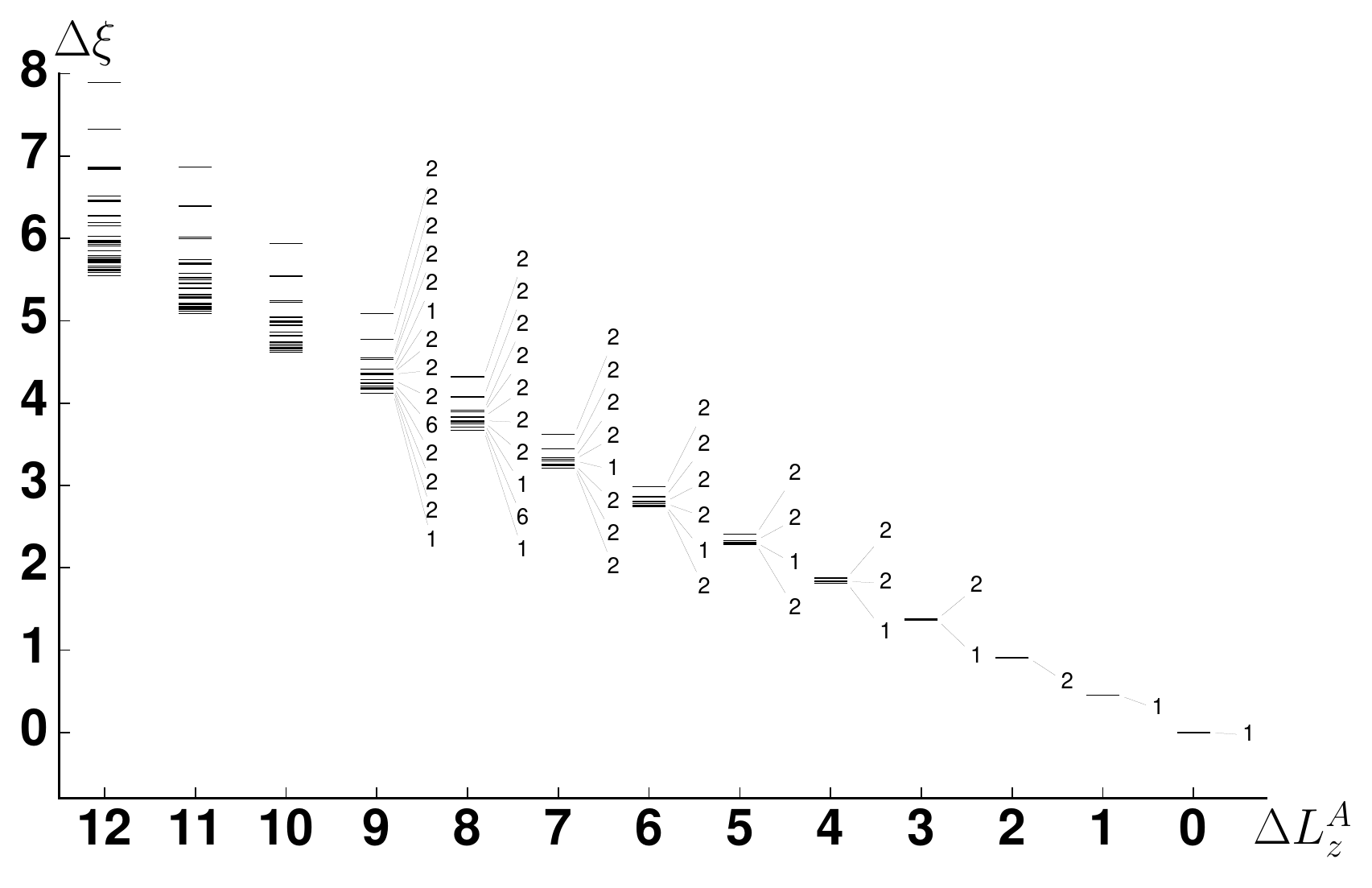}
%% (with legend turned off) 
%./fit_entanglement_spectrum.py --plot-rses --wf-type iqhe --nbr 50 --nbr-a 25 --jastrow 0 --rses-method exact --plot-sector 12 --rses-norm-zero --plot-degeneracy
\put(85,55){(a)}
\end{overpic}

\begin{overpic}[width=0.5\textwidth]{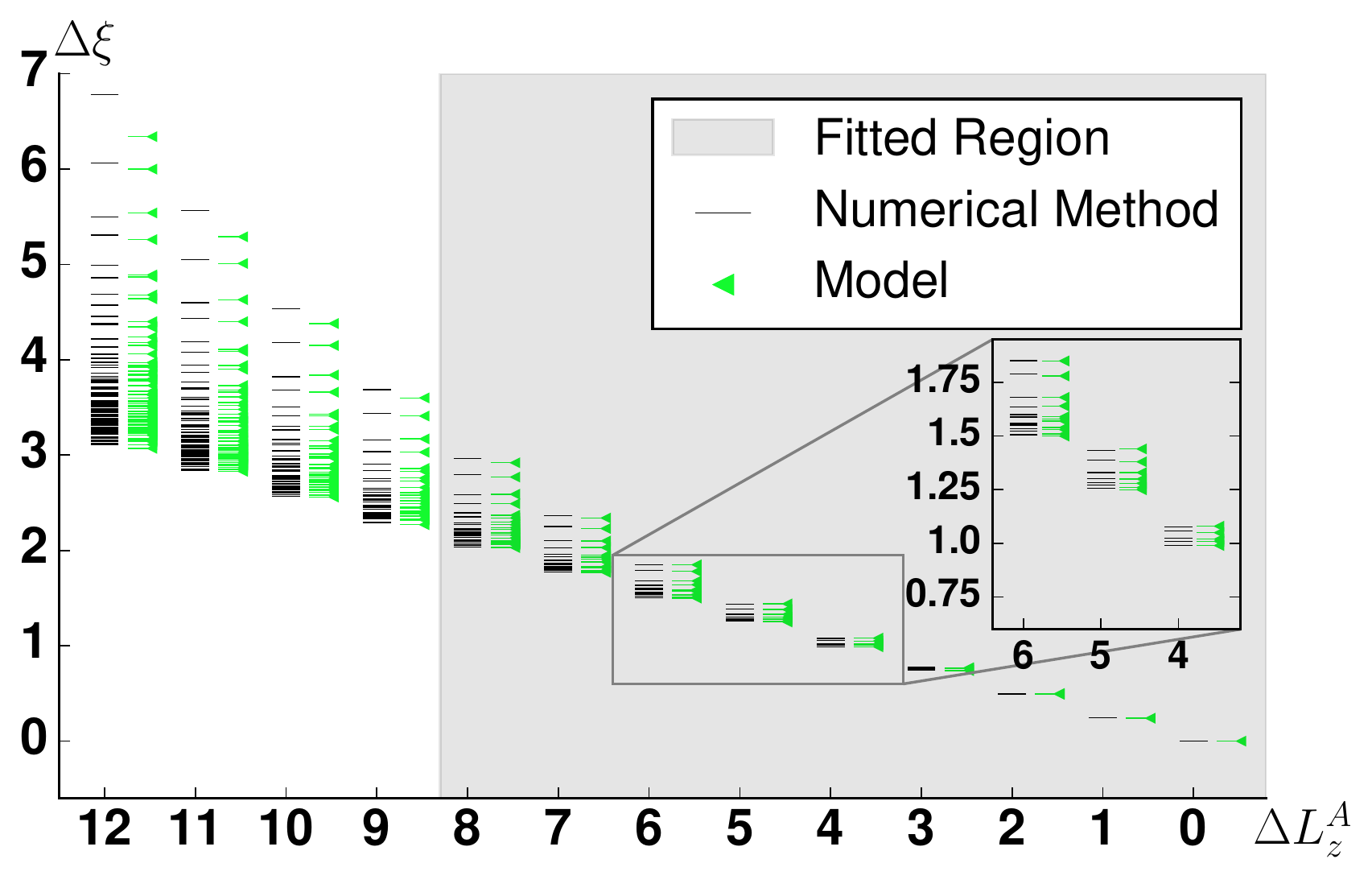}
%% (with ``fitting_model_processing.py'' edited to fit the 3 coefficients a_1,a_2,a_3)

%% ./fit_entanglement_spectrum.py --fit-rses --wf-type laughlin --nbr 50 --nbr-a 25 --nbr-ll 1 --jastrow 1 --fitting-model entanglement_energy --statistics bosons --rses-method ewf --rses-steps 6400 --rses-norm-zero --min-fit-sector 8 --max-fit-sector 8 

%% ./fit_entanglement_spectrum.py --make-plots --wf-type laughlin --nbr 50 --nbr-a 25 --nbr-ll 1 --jastrow 1 --fitting-model entanglement_energy --statistics bosons --rses-method ewf --rses-steps 6400 --rses-norm-zero --min-fit-sector 8 --max-fit-sector 8 --plot-sector 12 --plot-legend
\put(85,53){(b)}
\end{overpic}

\caption{
(color online). (a). Exact RSES for the $\nu=1$ state for $N=50$ particles, with equal-size $A$ and $B$ regions and $N_A=25$.  Embedded number labels indicate the degeneracy of $\xi$ (up to $\Delta L_z^A =8$). \cite{note1} (b). Fit of the single-particle entanglement energy model  (\eq{eqSingleParticleEnergyExpansion1} in terms of the composite fermion angular momentum $m'$ and truncated at order $m'^3$) to the ES of the $\nu=1/2$ Laughlin state calculated numerically for the same cut and system size using the method in \refr{rodriguez2013}. Fitting parameter data is included in \app{appFittingParameterData}. Entanglement energies $\xi$ and angular momentum labels $L_z^A$ are relative to the lowest lying state $\xi_0$ in the lowest angular momentum sector $L^A_0$ i.e. $\Delta \xi = \xi - \xi_0$ and $\Delta L_z^A = L_0^A - L^A_z$. 
}
\label{figLaughlinFit}
\end{figure}

\begin{figure}[t]

\begin{overpic}[width=0.5\textwidth]{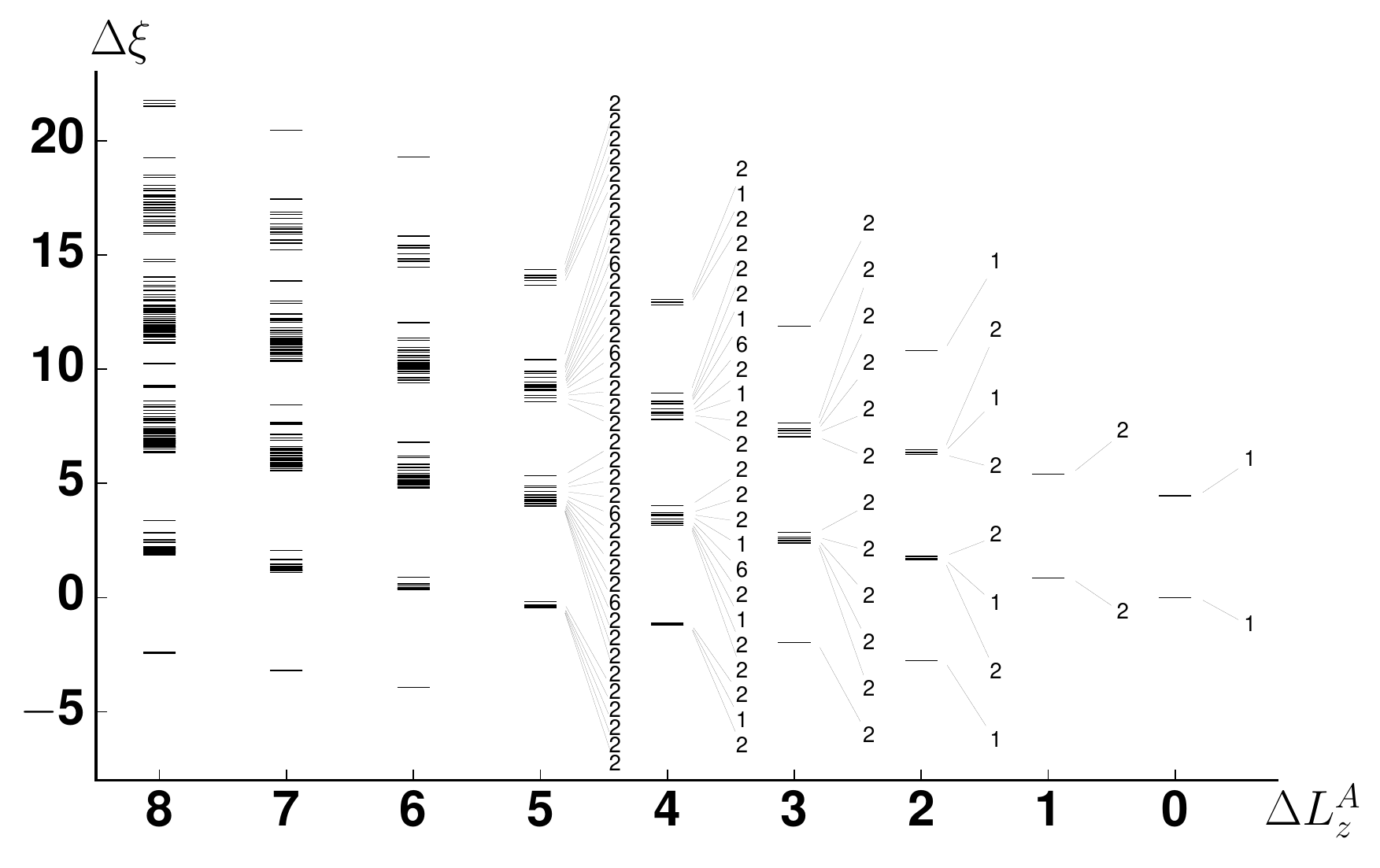}
%% ./fit_entanglement_spectrum.py --plot-rses --wf-type iqhe --nbr 48 --nbr-a 24 --jastrow 0 --rses-method exact --plot-sector 8 --rses-norm-zero --nbr-l 2 --plot-degeneracy
\put(85,53){(a)}
\end{overpic}

\begin{overpic}[width=0.5\textwidth]{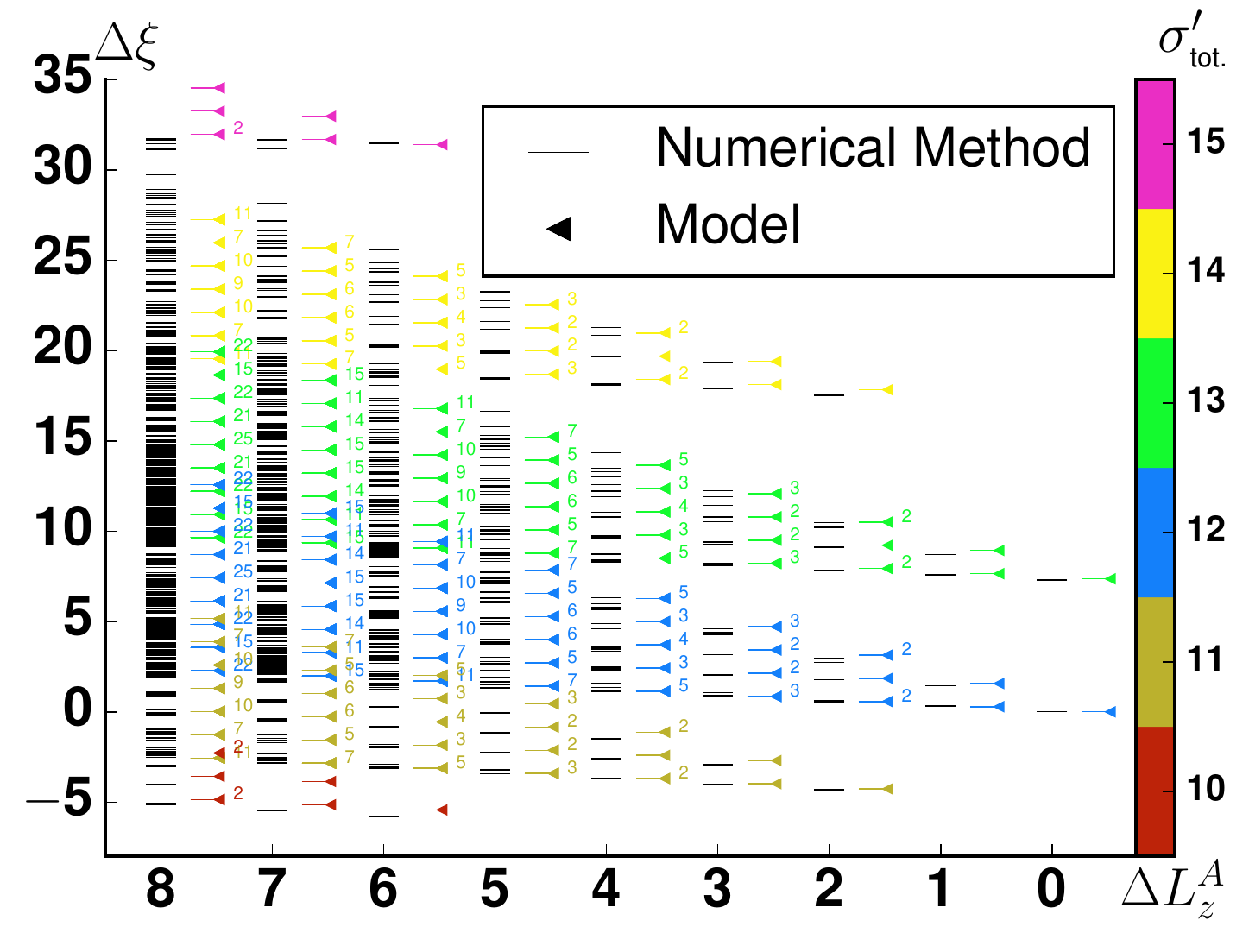}
%% (with ``fitting_model_processing.py'' edited with best-fit values of c0, b_0,a_1,b_1,. Also need to set initial b_0 to around +50)
%% ./fit_entanglement_spectrum.py --fit-rses --wf-type pef --nbr 48 --nbr-a 24 --nbr-ll 2 --jastrow 1 --fitting-model entanglement_energy --statistics bosons --rses-method ewf --rses-steps 6400 --rses-norm-zero --min-fit-sector 8 --max-fit-sector 8 --ewf-rses-path ewf_spectrum_unnormalized/

%% ./fit_entanglement_spectrum.py --make-plots --plot-sector 8 --wf-type pef --nbr 48 --nbr-a 24 --nbr-ll 2 --jastrow 1 --fitting-model entanglement_energy --statistics bosons --rses-method ewf --rses-steps 6400 --rses-norm-zero --min-fit-sector 8 --max-fit-sector 8 --ewf-rses-path ewf_spectrum_unnormalized/ --plot-legend --plot-degeneracy

\put(85,70){(b)}
\end{overpic}

\caption{
(color online). (a). Exact RSES for the $\nu=2$ state for $N=48$ particles, with equal-size $A$ and $B$ regions and $N_A=24$. (b). Fit of the single-particle entanglement energy model (\eq{eqSingleParticleEnergyJain}) to the $\nu=2/3$ Jain state calculated numerically for the same cut and system size using the method in \refr{rodriguez2013}. Fitting parameter data is included in \app{appFittingParameterData}. $\sigma'_{\mbox{\tiny tot.}}$ labels the number of particles in the upper effective Landau level for each $\xi$. Entanglement energies $\xi$ and angular momentum labels $L_z^A$ are relative to the lowest lying state $\xi_0$ in the lowest angular momentum sector $L^A_0$ i.e. $\Delta \xi = \xi - \xi_0$ and $\Delta L_z^A = L_0^A - L^A_z$. Number labels embedded in the plots indicate the degeneracy of $\xi$ where ambiguous. \cite{note1} 
}
\label{figJainFit}
\end{figure}

One can find exact expressions for the entanglement energies and Schmidt states in the case of a non-interacting Landau level problem,\cite{rodriguez2009}, using an argument proposed by Peschel. \cite{peschel2003} If one constructs the correlation matrix,
\begin{align}
C_{m_1,\sigma_1,m_2,\sigma_2} &= \int_{A} \phi^{*}_{m_1,\sigma_1}(\mathbf{r})\phi_{m_2,\sigma_2}(\mathbf{r})\, d\mathbf{r} \nonumber \\
&=\delta_{m_{1},m_{2}} \int_{A} \phi^{*}_{m_1,\sigma_1}(\mathbf{r})\phi_{m_1,\sigma_2}(\mathbf{r})\, d\mathbf{r}
\end{align}
then the eigenvalues, $\lambda_{m,\sigma}$, of $C$ are related to the entanglement energies by
\begin{equation}
\label{eqEntanglementEnergy}
\xi_{i} = \sum_{m,\sigma} o_{m,\sigma,i}\epsilon_{m,\sigma} + \mathrm{constant},
\end{equation}
where the \emph{single-particle entanglement energy function}, $\epsilon_{m,\sigma}$, is defined by
\begin{equation}
\epsilon_{m,\sigma}  = \mbox{log} \left[ \frac{1-\lambda_{m,\sigma}}{ \lambda_{m,\sigma} }\right].
\label{eqSingleParticleEntanglementEnergy}
\end{equation}
Note that now $\sigma$ labels the eigenstates of the correlation matrix at given $m$, rather than a Landau level. The full Schmidt states are Slater determinants built from these eigenstates and $o_{m,\sigma,i}$ denotes the occupation number of the correlation matrix eigenstates labelled by $(m,\sigma)$ in the Schmidt state with label $i$. Examples of the RSES for the $\nu=1$ and $\nu=2$ IQH states calculated using this method are included in \fig{figLaughlinFit}a and \fig{figJainFit}a for later comparison with our results for the FQH case. 

We now come to the statement of our main result: For the class of FQH states proposed by Laughlin and Jain to describe the FQH effect in the lowest Landau level at filling factors $\nu=n'/(2n'+1)$ [for fermions] and  $\nu=n'/(n'+1)$ [for bosons] with $n'$ integer, the low-lying $\xi$ in the associated RSES can be accurately described by the \emph{same} model as the IQH states (\eq{eqEntanglementEnergy}), but a \emph{different} single-particle entanglement energy function (\eq{eqSingleParticleEntanglementEnergy}). In addition, when multiple effective Landau levels are present in the Jain wave functions, one also needs to take into account a simple exchange-entanglement-energy term that we shall shortly describe with an example.

In this connection the Landau level index $\sigma$ for the IQH wave functions becomes the effective Landau level index of the Jain states, $\sigma'=0,1,\ldots,n'-1$, the magnetic field described by the monopole strength $Q$ is replaced with an effective magnetic field and an effective monopole strength $Q'= (N-n'^2)/n'$ and the occupation basis of single-particle Landau level orbitals now becomes an occupation basis of so-called composite fermion orbitals in the effective Landau levels, whose single-particle eigenstates are labelled by $m'=-(2Q'+\sigma')/2,-(2Q'+\sigma')/2+1,\ldots,(2Q'+\sigma')/2$. 

%%%%%%  NU=1 AND LAUGHLIN NU=1/2 FITS %%%%%%
%%%%%%  NU=1 AND LAUGHLIN NU=1/2 FITS %%%%%%

It is useful to expand $\epsilon_{m,\sigma}$ from \eq{eqSingleParticleEntanglementEnergy} as a power series in $m$ about the mid-point $m=0$, as it turns out that we need to keep only very few terms in this expansion for a very accurate description of $\epsilon_{m,\sigma}$. For example, in the case of a single Landau level ($\nu=\nll=1$ for spinless particles) or a single effective Landau level ($n'=1$ and replacing $m$ with $m'$ in the expansion)
\begin{equation}
\epsilon_{m,\sigma}  \approx a_0 + a_1 m + a_2 m^2 + a_3 m^3 + \ldots,
\label{eqSingleParticleEnergyExpansion1}
\end{equation}
where, for the IQH case, $a_0$, $a_1$ etc. are functions that can be expressed in terms of the eigenvalues of the matrix $C$ and its derivatives (note that they depend on the system size, via their dependence on the monopole strength $Q$). For 2 filled Landau levels ($\nu=\nll=2$ for spinless particles) or two effective Landau levels ($n'=2$) the expansion becomes instead
\begin{align}
\label{eqSingleParticleEnergyExpansion2}
\epsilon_{m,\sigma}  &  \approx a_{0,0}+ a_{0,1} m + a_{0,2} m^2 +a_{0,3} m^3 + \ldots  \\&+ [\sigma-1/2] \left\{a_{1,0} + a_{1,1} m + a_{1,2} m^2 +\ldots \right\}. \nonumber
\end{align}
In our approach for the FQH case the coefficients $a_{j,k}$ now play the role of fitting parameters in a model given by truncating the single particle energy function at a low degree in $m'$. Their values are determined by fitting this model to the RSES generated numerically from a microscopic construction of the FQH state using other methods. Further details of the fitting procedure are given in \app{appFittingAlgorithm}. To obtain an appropriate numerical spectrum to fit to for large system sizes we use the method proposed in \refr{rodriguez2013} (a method that works more efficiently for bosonic rather than fermionic states).

To illustrate how our approach works we present two examples of applications to bosonic FQH states, where we choose symmetrical, equal-sized regions $A$ and $B$ (i.e. a cut along the equator of the sphere) and $N_A=N_B=N/2$. \cite{note1} The method also applies for more general cuts. 

As our first example, we consider at the RSES of the Laughlin state of bosons at filling factor $\nu=1/2$ (this state also occurs in the Jain series for $n'=1$). The RSES for $\nu=1/2$ is accurately described by the single-particle energy function in \eq{eqSingleParticleEnergyExpansion1}, written in terms of the composite fermion angular momentum labels $m'$ and truncating the expansion at order $m'^3$. In this truncated expansion, the coefficients $a_1,a_2$ and $a_3$ are free fitting parameters. It is not necessary to include a constant term ($a_0$) in this fit since its value can be fixed by normalizing the spectrum (see \app{appFittingAlgorithm}).  An example fit to the numerically-calculated RSES of the $\nu=1/2$ state is shown in \fig{figLaughlinFit}b. Fitting parameter data is included in \app{appFittingParameterData}.  We find that by far the most dominant contribution comes from the $a_1$ term, indicating that the spectrum is almost linear. In \fig{figLaughlinFit}a we also show  the RSES of the IQH state for $\nu=1$ for comparison. 

The Laughlin state falls into the class of simple FQH states and its RSES has been studied previously using conformal field theory techniques. In particular \refr{dubail2012b} describes in detail a method to model the Laughlin RSES in terms of the energy levels of an entanglement Hamiltonian that is given by writing down every allowable conformal field theory operator (in this case chiral boson operators, their derivatives and powers) order-by-order in their scaling dimension (with terms at lower scaling dimension providing more relevant contributions). These operators come with unknown coefficients that can be fitted to the numerically-calculated RSES using a similar procedure to that described in \app{appFittingAlgorithm}. In \refr{dubail2012b}  it was shown that the RSES of the $\nu=1/2$ state can be very accurately described by truncating at scaling dimension 2, leading to an entanglement Hamiltonian containing 3 terms (and therefore 3 unknown fitting parameters). We have checked that for the same data set as used here (\fig{figLaughlinFit}b) the fit of this truncated entanglement Hamiltonian to the RSES is in excellent agreement with our non-interacting model (in fact, the quality of the fit as defined in \app{appFittingAlgorithm} is typically improved by an order of magnitude compared to our approach). One could in principle reproduce our model directly (with the inclusion of additional interaction terms) by fermionizing Dubail--Read--Rezayi's entanglement Hamiltonian. The fact that our result agrees closely with the numerically-calculated RSES implies that any additional interaction terms must only be small corrections, which justifies the assumptions made by the composite fermion approach. 

%%%%%%  NU=2 AND JAIN NU=2/3 FITS %%%%%%
%%%%%%  NU=2 AND JAIN NU=2/3 FITS %%%%%%

Our second example is the RSES of the Jain state of bosons at filling factor $\nu=2/3$ (this state occurs in the Jain series for $n'=2$ for bosons). Our technique was not able to improve upon conformal field theory in treating simple FQH wave functions, but where it really excels is in its treatment of the Jain states.  In this case we now use the single-particle energy from the $\nu=2$ IQH state (\eq{eqSingleParticleEnergyExpansion2}), but we augment it to take into account an exchange-like interaction (treated at the level of ``mean-field theory'') for fermions in different Landau levels. This term can alternatively be thought of as a ``charging energy''. Our ansatz to describe the RSES of the $\nu=2/3$ Jain state is
\begin{equation}
\xi_{N_A,L_z^A ,i} = \sum\limits_{m,\sigma} n_{m,\sigma} \epsilon_{m,\sigma} + c \left(\Delta N \right )^2,
\label{eqExtraTerm}
\end{equation}
with the single-particle energy function given by
\begin{equation}
\epsilon_{m',\sigma'}  \approx a_{0,1} m' + [\sigma' -1/2] \left\{a_{1,0} + a_{1,1} m'\right\},
\label{eqSingleParticleEnergyJain}
\end{equation}
and the exchange-interaction, or charging energy term, given by 
\begin{equation}
\Delta N = \sum\limits_{m',\sigma'} n_{m',\sigma'}   [\sigma' -1/2].
\label{eqExchangeInteraction}
\end{equation}
This ansatz requires 4 fitting parameters $a_{0,1},a_{1,0},a_{1,1}$ and $c$ (once again, a constant term $a_{0,0}$ is not fitted because it can be fixed by normalization).

In \fig{figJainFit}b we show an example fit to the RSES of the $2/3$ state for a large system of $N=48$ particles. Fitting parameter data is included in \app{appFittingParameterData}. We observe that the most dominant contribution arises from the coefficient $a_{1,0}$, which can be thought of as an entanglement ``cyclotron energy'' term. We also find that the coefficient of the exchange interaction, $c$, is positive (so due to this additional term, the branches in the RSES for $\nu=2/3$ are further apart than they would have been otherwise). Note that there are degeneracies in the fitted entanglement energy eigenvalues, but these can be lifted by truncating the entanglement energy expansion at higher order in $m'$. For comparison, in \fig{figJainFit}a we plot the RSES of the $\nu=2$ IQH state. 

%%%%%%	CONCLUSION	%%%%%%%%%%%%%%%%%%%%%%

To summarize, we have described how a RSES of non-interacting composite fermions can be constructed to accurately approximate the RSES for certain strongly correlated FQH states. Key to this construction is the observation that the single-particle entanglement energy function underlying the description of the RSES for the IQH states can be simply modified in order to describe the RSES of FQH states, treating the many-body RSES within a non-interacting approximation and allowing for very basic mean-field theory like exchange energy corrections in the multi-Landau level case. We note that this description of the RSES closely parallels the description of the real  energy spectrum of the edge of a Hall droplet in terms of composite fermions (see e.g. \refr{Sreejith2011}).

The quality of this approximation could be improved by allowing for higher order corrections to the underlying entanglement energy function (at the cost of needing more fitting parameters). These corrections can be thought of as allowing for additional inter-composite fermion interactions at the level of ``mean field theory''. Alternatively, weak interaction corrections can be added to the model perturbatively, working in the framework of degenerate perturbation theory as applied to the entanglement Hamiltonian. However, we find that such corrections only provide a marginal improvement to the fit. 

\newpage

%%%%%%  ACKNOWLEDGEMENTS  %%%%%%%%%%%%%%
\vspace{0.5em}
\noindent \textbf{Acknowledgements:} We thank S.~Sondhi for helpful discussions. SCD was supported by EPSRC grant EP/J017639/1. IDR was supported by EU project SIQS. JKS was supported by Science Foundation Ireland Principal Investigator Award 12/IA/1697. SHS was supported by EPSRC grants EP/I032487/1 and EP/I031014/1. We acknowledge use of the Hydra computer cluster at the Rudolf Peierls Centre for Theoretical Physics. Statement of compliance with EPSRC policy framework on research data: This publication reports theoretical work that does not require supporting research data.

%%%%%%  BIBLIOGRAPHY  %%%%%%%%%%%%%%%%%%

\bibliographystyle{prsty}
\bibliography{myrefs}

\begin{thebibliography}{10}

\bibitem{li2008}
H. Li and F.~D.~M. Haldane, Phys. Rev. Lett. {\bf 101},  010504  (2008).

\bibitem{haque2007}
M. Haque, O. Zozulya, and K. Schoutens, Phys. Rev. Lett. {\bf 98},  060401
  (2007).

\bibitem{zozulya2007}
O.~S. Zozulya, M. Haque, K. Schoutens, and E.~H. Rezayi, Phys. Rev. B {\bf 76},
   125310  (2007).

\bibitem{dubail2012a}
J. Dubail, N. Read, and E.~H. Rezayi, Phys. Rev. B {\bf 85},  115321  (2012).

\bibitem{sterdyniak2012}
A. Sterdyniak, A. Chandran, N. Regnault, B.~A. Bernevig, and P. Bonderson,
  Phys. Rev. B {\bf 85},  125308  (2012).

\bibitem{rodriguez2012}
I.~D. Rodr\'{i}guez, S.~H. Simon, and J.~K. Slingerland, Phys. Rev. Lett. {\bf
  108},  256806  (2012).

\bibitem{qi2012}
X.-L. Qi, H. Katsura, and A.~W.~W. Ludwig, Phys. Rev. Lett. {\bf 108},  196402
  (2012).

\bibitem{swingle2012}
B. Swingle and T. Senthil, Phys. Rev. B {\bf 86},  045117  (2012).

\bibitem{dubail2012b}
J. Dubail, N. Read, and E.~H. Rezayi, Phys. Rev. B {\bf 86},  245310  (2012).

\bibitem{lopez1991}
A. Lopez and E. Fradkin, Phys. Rev. B {\bf 44},  5246  (1991).

\bibitem{moore1991}
G. Moore and N. Read, Nucl. Phys. {B} {\bf 360},  363  (1991).

\bibitem{wen1992}
X.-G. Wen, International Journal of Modern Physics B {\bf 06},  1711  (1992).

\bibitem{cappelli1993}
A. Cappelli, C.~A. Trugenberger, and G.~R. Zemba, Nuclear Physics B {\bf 396},
  465   (1993).

\bibitem{jainbook}
J. K. Jain, {\it Composite Fermions}, (Cambridge University Press, Cambridge,
  2007).

\bibitem{hansson2007a}
T.~H. Hansson, C.-C. Chang, J.~K. Jain, and S. Viefers, Phys. Rev. Lett. {\bf
  98},  076801  (2007).

\bibitem{hansson2007b}
T.~H. Hansson, C.-C. Chang, J.~K. Jain, and S. Viefers, Phys. Rev. B {\bf 76},
  075347  (2007).

\bibitem{bergholtz2008}
E.~J. Bergholtz, T.~H. Hansson, M. Hermanns, A. Karlhede, and S. Viefers, Phys.
  Rev. B {\bf 77},  165325  (2008).

\bibitem{cappelli2013}
A. Cappelli, Journal of Physics A: Mathematical and Theoretical {\bf 46},
  012001  (2013).

\bibitem{peschel2003}
I. Peschel, Journal of Physics A: Mathematical and General {\bf 36},  L205
  (2003).

\bibitem{rodriguez2009}
I.~D. Rodr\'iguez and G. Sierra, Phys. Rev. B {\bf 80},  153303  (2009).

\bibitem{rodriguez2010}
I.~D. Rodr\'iguez and G. Sierra, Journal of Statistical Mechanics: Theory and
  Experiment {\bf 2010},  P12033  (2010).

\bibitem{haldane1983}
F.~D.~M. Haldane, Phys. Rev. Lett. {\bf 51},  605  (1983).

\bibitem{wu1976}
T.~T. Wu and C.~N. Yang, Nuclear Physics B {\bf 107},  365   (1976).

\bibitem{note1}
Note that due to our choice of equal-sized regions $A$ and $B$, degeneracies
  occur in the RSES of the integer quantum Hall states. These degeneracies can
  be associated with a particle-hole symmetry in the occupations of the
  single-particle orbitals in the Schmidt decomposition, \eq{eqSchmidt}. Such
  particle-hole degeneracy is no longer present once interactions are
  introduced as can be seen in e.g. \fig{figLaughlinFit}b.

\bibitem{rodriguez2013}
I.~D. Rodr\'{i}guez, S.~C. Davenport, S.~H. Simon, and J.~K. Slingerland, Phys.
  Rev. B {\bf 88},  155307  (2013).

\bibitem{Sreejith2011}
G.~J. Sreejith, S. Jolad, D. Sen, and J.~K. Jain, Phys. Rev. B {\bf 84},
  245104  (2011).

\end{thebibliography}

%%%%%%  APPENDIX  %%%%%%%%%%%%%%%%%%

\appendix

\section{Fitting algorithm}
\label{appFittingAlgorithm}

In this appendix we shall briefly describe our procedure for evaluating the goodness of fit of the RSES to the single-particle models proposed in this work. 

The problem is a many-parameter optimization where we aim to minimize a weighted sum of squared differences between corresponding $\xi$ in the lowest lying part of the RSES and single-particle energy spectrum for a given set of fitting parameters $\left\{a_{j,k}\right\}$. By corresponding $\xi$ we mean that, for comparison, we normalize both the model spectrum and the numerically calculated spectrum such that the lowest lying $\xi$ in the lowest $\Delta L_z^A$ sector is set to zero. Then we order the set of $\xi$ in the model spectrum (call them $\xi_{\mbox{\tiny model}}$) and the lowest lying part of the RSES (call them $\xi_{\mbox{\tiny RSES}}$) by their $\Delta L_z^A$ values, then for each sector we sort in order of increasing $\xi_{\mbox{\tiny RSES}}$ or $\xi_{\mbox{\tiny model}}$ value and finally we take the sum of squared difference between the lowest $\xi_{\mbox{\tiny RSES}}$ and lowest $\xi_{\mbox{\tiny model}}$, the difference between the next lowest $\xi_{\mbox{\tiny RSES}}$ and next lowest $\xi_{\mbox{\tiny model}}$ and so on. This sum can also be weighted in various ways, for instance if we want to give increased importance to matching up states with the lowest values of $\xi$. In general, therefore, we aim to minimize a fitting function of the form
\begin{equation}
\label{eqFittingParameterDefinition}
R(\left\{a_{j,k}\right\}) = \sum_i \left[\xi^i_{\mbox{\tiny RSES}} -\xi^i_{\mbox{\tiny model}}(\left\{a_{j,k}\right\})\right]^2 W_i,
\end{equation}
where $\left\{a_{j,k}\right\}$ denotes the set of free parameters in the model, and the index $i$ appearing in the sum denotes the $i$th value of $\xi_{\mbox{\tiny model}}$, $\xi_{\mbox{\tiny RSES}}$ within the ordered set (ordered in the sense described above). The factor $W_i$ assigns an optional weight to each term in the sum. In addition, we only include entanglement spectrum eigenvalues below a certain specified cut-off sector $\Delta L^A_{\mbox{\tiny cut-off}}$, i.e. $W_i=0$ if it refers a state with $\Delta L_z^A > \Delta L^A_{\mbox{\tiny cut-off}}$. 

In our application we always include a factor in $W_i$ that divides out the number of states in each sector (i.e. for every $\xi^i_{\mbox{\tiny EH}}$ labelled by the same $\Delta L_z^A$ we divide by a factor of $N_{\Delta L_z^A}$ that counts the total number of $\xi_{\mbox{\tiny model}}$ with that $\Delta L_z^A$). The reason for doing this is because otherwise the fitting function would assign overwhelmingly more weight to fitting the higher sectors (because the counting of states in the RSES grows superpolynomially). Once the $N_{\Delta L_z^A}$ factor is divided out, each sector counts for the same total weight in \eq{eqFittingParameterDefinition}, i.e. $\sum_{i \mbox{\tiny\,with\,} \Delta L_z^A}  W_i =1$ for all $\Delta L_z^A$. Another aspect to consider is that, in the RSES, linearly smaller $\xi$ correspond to exponentially greater coefficients in the Schmidt decomposition. In this sense, it should be more important physically to match up the lower lying $\xi$. For this reason, we might also consider including a factor $\mbox{exp}(-\xi^i_{\mbox{\tiny RSES}})$ in our weight function. This addition works well for the Laughlin case, however the Jain case is more complicated due to the presence of the branch structure in the RSES. Consequently, for studying the Laughlin state we use the fitting function
\begin{equation}
R_{\mbox{\tiny Laughlin}} = \sum\limits_{i \, \mid \,  \Delta L_z^A \le \Delta L^A_{\mbox{\tiny cut-off}}} \frac{\left(\xi^i_{\mbox{\tiny RSES}} -\xi^i_{\mbox{\tiny model}}\right)^2 e^{-\xi^i_{\mbox{\tiny RSES}}}}{N_{\Delta L_z^A}},
\label{eqFitParameterLaughlin}
\end{equation}
whereas for fitting the Jain state we find that for fitting all branches with equal weight it is better to use
\begin{equation}
R_{\mbox{\tiny Jain}} = \sum_{i \, \mid \, \Delta L_z^A \le \Delta L^A_{\mbox{\tiny cut-off}}} \frac{\left(\xi^i_{\mbox{\tiny RSES}} -\xi^i_{\mbox{\tiny model}}\right)^2 }{N_{\Delta L_z^A}}.
\label{eqFitParameterJain}
\end{equation}

The quality of a given fitting model (e.g. a given set of fitting parameters $\left\{a_{j,k}\right\}$) can be assessed by the minimal value of the fitting function obtained for that model. The lower the minimal value of the fitting function, the better the quality of the fit. 

In order to solve the minimization problem we use the Powell method provided by Scientific Python (SciPy) version 0.11 and above, which involves a sequential 1D minimization of each fitting parameter. The Powell method is found to be numerically stable for this problem. One also has to take steps to avoid finding the local rather than the global minimum, which we achieve by running the procedure a large number of times with different random starting parameters. 

\section{Fitting parameter data}
\label{appFittingParameterData}

In this appendix we tabulate the values of the fitting parameters $a_{j,k}$ obtained in the single-particle energy fits shown in \fig{figLaughlinFit} and \fig{figJainFit}. We also include the corresponding values of the fitting function $R(\left\{a_{j,k}\right\})$ defined in \app{appFittingAlgorithm} to assess the relative quality of those fits. 

Example fitting parameter values and  $R_{\mbox{\tiny Laughlin}}$  (\eq{eqFitParameterLaughlin}) values in the single-particle model proposed in this work for the Laughlin $\nu=1/2$ state. Corresponding results plotted in \fig{figLaughlinFit}. Smaller values of $R_{\mbox{\tiny Laughlin}}$ indicate a better quality of fit:

\begin{minipage}{\columnwidth}
\vspace{1em}
\begin{center}
\begin{tabularx}{\columnwidth}{ *{5}{R} }             
  $a_1$ &  $a_2$ & $ a_3$ & $\Delta L^A_{\mbox{\tiny cut-off}}$ & $R_{\mbox{\tiny Laughlin}}$ \\
    \hline               
  $0.243$  & $0.00130$  &  $0.00214$ & 8 & $8.12\times10^{-5}$
\end{tabularx}
\end{center}
\vspace{1em}
\end{minipage}

Example fitting parameter values and  $R_{\mbox{\tiny Jain}}$ (\eq{eqFitParameterJain}) values in the single-particle model proposed in this work for the Jain $\nu=2/3$ state. Corresponding results plotted in \fig{figJainFit}. Smaller values of $R_{\mbox{\tiny Jain}}$ indicate a better quality of fit:

\begin{minipage}{\columnwidth}
\vspace{1em}
\begin{center}
\begin{tabularx}{\columnwidth}{ *{6}{R} }
  $a_{0,1}$ & $a_{1,0}$ & $a_{1,1}$ & $c$ & $\Delta L^A_{\mbox{\tiny cut-off}}$ & $R_{\mbox{\tiny Jain}} $ \\
    \hline
   $0.9279$ &  $3.371$ & $0.6429$ & $0.1557$ & 8 & $ 0.945$\\

\end{tabularx}
\end{center}
\vspace{1em}
\end{minipage}

\end{document}